# Maximizing the spin-orbit torque efficiency of Pt/Ti multilayers by optimization of the tradeoff between the intrinsic spin Hall conductivity and carrier lifetime

Lijun Zhu* and R. A. Buhrman
*Cornell University, Ithaca, New York 14850, USA*

*lz442@cornell.edu

We report a comprehensive study of the maximization of the spin Hall ratio ($\theta_{SH}$) in Pt thin films by the insertion of sub-monolayer layers of Ti to decrease carrier lifetime while minimizing the concurrent reduction in the spin Hall conductivity. We establish that the intrinsic spin Hall conductivity of Pt, while robust against the strain and the moderate interruption of crystal order caused by these insertions, begins to decrease rapidly at high resistivity level because of the shortening carrier lifetime. The *unavoidable* trade-off between the intrinsic spin Hall conductivity and carrier lifetime sets a practical upper bound of $\theta_{SH} \geq 0.8$ for heterogeneous materials where the crystalline Pt component is the source of the spin Hall effect and the resistivity is increased by shortening carrier lifetime. This work also establishes a very promising spin-Hall metal of [Pt 0.75 nm/Ti 0.2 nm]$_7$/Pt 0.75 nm for energy-efficient, high-endurance spin-orbit torque technologies (e.g., memories, oscillators, and logic) due to its combination of a giant $\theta_{SH} \approx 0.8$, or equivalently a dampinglike spin torque efficiency per unit current density $\xi_{DL}^j \approx 0.35$, with a relatively low resistivity (90 μΩ cm) and high suitability for practical technology integration.

Spin Hall metals with strong dampinglike spin-orbit torque (SOT) efficiency per unit current density ($\xi_{DL}^j$) and relatively low resistivity ($\rho_{xx}$) are the key for developing practical spin-orbit torque technologies (e.g., memories, oscillators, and logic)[1-10] that require energy efficiency, high endurance, and low impedance [11,12]. For example, for a SOT-MRAM device with a spin Hall channel (with thickness $d_{HM}$ and resistivity $\rho_{xx}$)/FM free layer (with thickness $t$ and resistivity $\rho_{FM}$), the write current is approximately $I_{write} \propto (1+s)/\xi_{DL}^j$, and the corresponding write energy is $P_{write} \propto [(1+s)/\xi_{DL}^j]^2 \rho_{xx}$, where $s \approx t\rho_{xx}/d_{HM}\rho_{FM}$ is the ratio of the current shunting in the FM layer over the current flow in the spin Hall channel (see Fig. 1(a)). Meanwhile, a high $\rho_{xx}$ of a spin Hall material (e.g., $\rho_{xx} \geq 200$ μΩ cm in Ta [2], W [8,13,14], and topological insulators [15,16]) is problematic for applications that require a high endurance [17] and low device impedance ($\propto \rho_{xx}$)[18]. For example, use of a large-$\rho_{xx}$ spin Hall material (e.g., $\rho_{xx}$ =200 -300 μΩ cm for W, see Fig. 1(b)) will limit the endurance of SOT devices via Joule-heating-induced bursting and migration of the write line [17] as well as result in a high write impedance that is difficult for superconducting circuits in a cryogenic computing system to accommodate [18]. It is therefore of great technological and fundamental importance to establish how, why, and to what limit the spin Hall ratio ($\theta_{SH}$) and $\xi_{DL}^j$ of a spin Hall metal with a giant spin Hall conductivity ($\sigma_{SH}$) and a relatively low $\rho_{xx}$ can be enhanced in practice.

Among the various spin Hall metals, Pt is particularly attractive for spin-torque technologies due to its low $\rho_{xx}$, the highest intrinsic $\sigma_{SH}$ known for the simple technologically viable metals (> $1.6\times10^6$ ($\hbar/2e$) Ω$^{-1}$ m$^{-1}$ in the clean–metal regime)[11,12,19-24], easy growth, and the capability to be readily integrated into experimental and/or manufacturing processes. However, $\xi_{DL}^j$ =(2e/$\hbar$)$T_{int}\sigma_{SH}\rho_{xx}$ for low-$\rho_{xx}$ Pt films is considerably lower than that of the meta-stable $\beta$-W alternative [8,13,14] due to the much lower $\rho_{xx}$, e.g. $\xi_{DL}^j$= 0.06 when $\rho_{xx}$ = 20 μΩ cm [25]. Here, $e$ is the elemental charge, $\hbar$ the reduced Planck's constant, and $T_{int}$ the spin transparency of the HM/FM interface [26-31]. To take better advantage of the intrinsic nature and very high magnitude of $\sigma_{SH}$ of Pt [11,12,19-24], recent efforts have sought to enhance $\xi_{DL}^j$ by increasing $\rho_{xx}$ through the addition of a high density of impurities [22,23], by alloying [11,12], or by insertion of multiple sub-monolayers of a second material into a Pt film to take advantage of strong interfacial scattering [24]. However, raising $\rho_{xx}$ via any of these approaches eventually results in a rapid degradation of the dampinglike SOT efficiency per applied field $\xi_{DL}^E$=(2e/$\hbar$)$T_{int}\sigma_{SH}$ [22-24], which may set a practical limit to which $\xi_{DL}^j = \xi_{DL}^E \rho_{xx}$ can be enhanced via these approaches. Insights into the mechanisms of such degradation of $\xi_{DL}^E$ and $\sigma_{SH}$ are the key for developing new effective techniques to optimize $\xi_{DL}^j$ for high-performance spin-torque applications through raising $\rho_{xx}$ without *avoidable* degradation of $\sigma_{SH}$.

In this work, we report the evolution of the spin Hall effect (SHE) of Pt achieved with the progressive insertion of multiple sub-monolayer (0.2 nm) layers of Ti into the Pt films. These insertions resulted in increases in $\rho_{xx}$ due to strong interfacial scattering but did not materially degrade the basic face-centered cubic (fcc) order of Pt. We surmise that this latter is the result of the almost identical atomic radii of Ti and Pt. At the optimum density of Ti insertion layers, we achieved a maximum $\theta_{SH} \approx 0.8$ and $\xi_{DL}^j \approx 0.35$ as measured with a thin Co SOT detector layer, while increasing $\rho_{xx}$ from 26.5 (no Ti insertion layers) to 90 μΩ cm. Upon further increasing the insertion layer density $\rho_{xx}$ continued to increase quasi-linearly with insertion layer density, but $\xi_{DL}^j$ begins to slowly decrease due to a more rapid decrease in $\sigma_{SH}$. We establish that this unavoidable reduction of $\sigma_{SH}$ is mainly due to the effect of the shortening of carrier lifetime, while is insensitive to the strain and the moderate interruption of crystal order of Pt.

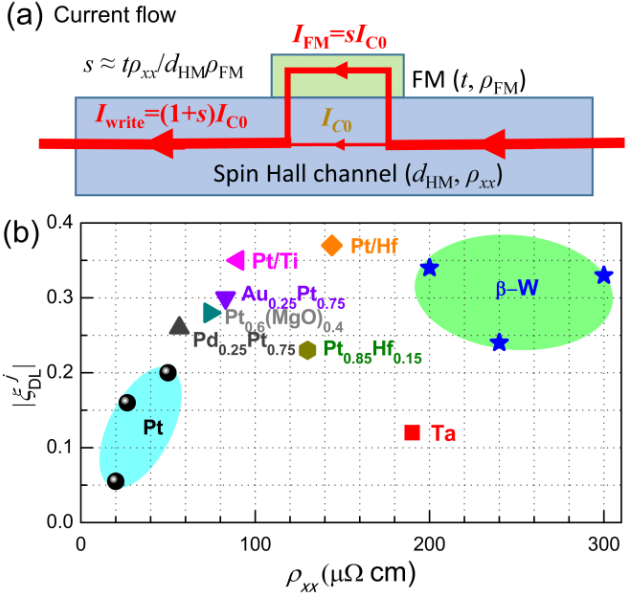

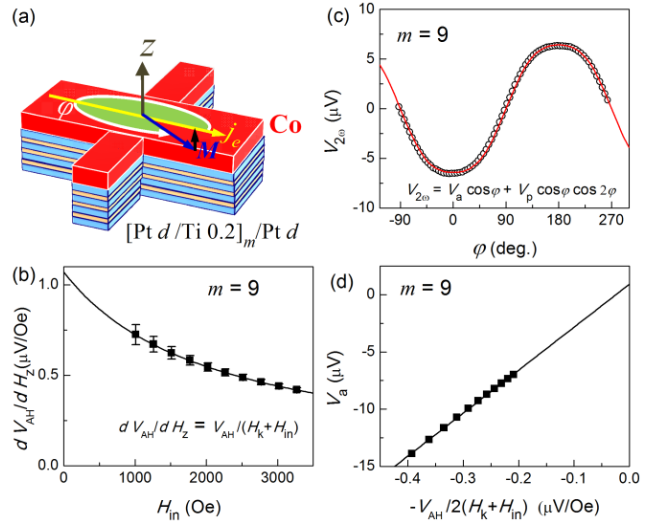

Fig. 1 (a) Cartoon of current flow in a spin Hall channel and a FM dot, the core structure of spin-orbit torque memory, logic and oscillator devices. $I_{c0}$ is the critical switching current in the spin Hall channel (thickness $d_{HM}$, resistivity $\rho_{xx}$) for SOT generation, while $I_{FM}= sI_{C0}$ represents the current shunting into the FM layer (thickness $t$, resistivity $\rho_{FM}$). (b) Comparison of $\xi_{DL}^j$ and $\rho_{xx}$ of HM/FM bilayers. The values of $\xi_{DL}^j$ and $\rho_{xx}$ are taken from Ref. [7,35] for Pt, Ref. [2] for $\beta$-Ta, Ref. [22] for $Pt_{0.85}Hf_{0.15}$, Ref. [24] for Pt/Hf multilayers, Ref. [8,13,14] for $\beta$-W, Ref. [12] for $Pd_{0.25}Pt_{0.75}$, Ref. [11] for $Au_{0.25}Pt_{0.75}$, and Ref. [23] for $Pt_{0.6}(MgO)_{0.4}$.

Fig. 2. (a) Schematic of the "in-plane" harmonic response measurement. (b) In-plane field ($H_{in}$) dependence of $dV_{AH}/dH_z$, (c) In-plane field angle dependence of the second harmonic voltages ($V_{2\omega}$)($H_{in}$ = 3 kOe), and (d) $V_a$ versus $-V_{AH}/2(H_{in}+H_k)$ for [Pt 0.6/Ti 0.2]$_9$/Pt 0.6/Co 1.3. The solid lines in (b)-(d) refer to the best fits of the data to $dV_{AH}/dH_z =V_{AH}/(H_{in}+H_k)$, $V_{2\omega}=V_a\cos\varphi+V_p\cos\varphi\cos2\varphi$, and $V_a= -H_{DL}V_{AH}/2(H_{in}+H_k) +V_{ANE}$, respectively.

For this study, we prepared magnetic stacks of [Pt $d$/Ti 0.2]$_m$/Pt $d$/Co $t$ (numbers are layer thicknesses in nm) with $(m+1)d = 6$ nm and $m = 0, 1, 3, 5, 7, 9$, and 11, respectively, Each sample was sputter-deposited onto Si/SiO$_2$ substrates with a 1 nm Ta seed layer and a capping layer of MgO 2.0 / Ta 1.5 bilayers. We chose $t = 1.9$ nm for $m = 0$ and 1.3 nm for $m \geq 1$ to assure that the Co layer is thick enough to be fully in-plane magnetized as well as being thin enough to have a strong dampinglike effective SOT field ($H_{DL}$) during the "in-plane" harmonics response measurement (Fig. 2(a)). As determined by vibrating sample magnetometry, the effective saturation magnetization ($M_s^{eff}$) of the as-grown Co films, the total moment averaged over the volume of the Co films, are 1180-1250 emu/cc, depending on the details of the sample interfaces. The samples were patterned into 5×60 μm$^2$ Hall bars by photolithography and ion milling, followed by deposition of 5 nm Ti and 150 nm Pt as contacts.

Within the macrospin approximation, the dampinglike SOT effective field ($H_{DL}$) for in-plane magnetized HM/FM samples can be determined from the angle ($\varphi$) dependence of the second harmonic response voltage ($V_{2\omega}$) on the in-plane bias magnetic field ($H_{in}$)[11,12], i.e. $V_{2\omega} = V_a\cos\varphi + V_p\cos\varphi\cos2\varphi$, where $V_a = -V_{AH}H_{DL}/2(H_{in}+H_k) + V_{ANE}$, with $V_{AH}$, $V_{ANE}$, and $H_k$ being the anomalous Hall voltage, the anomalous Nernst effect, and the perpendicular anisotropy field. $V_p$ term is the contribution of fieldlike and Oersted torques, respectively. In this work, we used a sinusoidal electric bias field with a constant magnitude of $E = 66.7$ kV/m as the excitation for the harmonic response measurement. As shown in Fig. 2(b), we first determined $V_{AH}$ and $H_k$ simultaneously by fitting the $H_{in}$ dependence of $dV_{AH}/dH_z$ to the relation $dV_{AH}/dH_z = V_{AH}/(H_{in}+H_k)$, where $dV_{AH}/dH_z$ is the the derivative of $V_{AH}$ with respect to the swept out-of-plane field ($H_z$) under different fixed in-plane bias field $H_{in}$. $V_a$ for each magnitude of $H_{in}$ was separated out from the $\varphi$ dependence of $V_{2\omega}$ (Fig. 2(c)). As shown in Fig. 2(d), the linear fit of $V_a$ vs $V_{AH}/2(H_{in}+H_k)$ gives the value of $H_{DL}$ of 5.6×10$^{-4}$ Oe m/V (the slope) and a negligible thermal effect ($V_{ANE} < 0.9$ μV from the intercept).

Using the values of $H_{DL}$, we determined the SOT efficiencies for the [Pt $d$/Ti 0.2]$_m$/Pt $d$/Co $t$ samples via $\xi_{DL}^E = (2e/\hbar)\mu_0 M_s^{eff} t H_{DL}/E$, where $\mu_0$ is the permeability of vacuum. Here we use $M_s^{eff}$ in the calculation of the SOT efficiencies to include any contribution of possible magnetic dead layers [29] and/or proximity magnetism [32] at the interfaces. The uncertainty of $\xi_{DL}^E$ due to the harmonic response measurement is less than 2%. As shown in Fig. 3(a), $\xi_{DL}^E$ drops monotonically with the Ti insertions from 6.1×10$^5$ Ω$^{-1}$ m$^{-1}$ for $m = 0$ ($d = 6$ nm) to 2.6×10$^5$ Ω$^{-1}$ m$^{-1}$ for $m = 11$ ($d = 0.5$ nm). Figure 3(b) shows the corresponding results for $\xi_{DL}^j = (2e/\hbar)\mu_0 M_s^{eff} t H_{DL}\rho_{xx}/E$, where $\rho_{xx}$ for each Pt/Ti multilayer sample was determined by measuring the resistance enhancement of the stack with $m$ insertions relative to the reference stack with $m = 0$. $\xi_{DL}^j$ increased quickly from ~0.16 at $d = 6$ nm ($m = 0$) to a peak value of ~0.35 at $m = 7$ ($d = 0.75$ nm) and then dropped slightly to 0.33 at $m = 11$ ($d = 0.5$ nm), while, as plotted in Fig. 3(c), $\rho_{xx}$ increased monotonically, from 26.6 μΩ cm for $m = 0$ to 192.3 μΩ cm for $m = 11$, due to the increased interfacial scattering added by each Ti insertion layer. To best understand the physics responsible for the evolution of $\xi_{DL}^{j(E)}$ with Ti insertions, we need to determine the bulk values of $\theta_{SH} = \xi_{DL}^{j}/T_{int}$ and $\sigma_{SH} = (\hbar/2e)\xi_{DL}^{E}/T_{int}$ for the Pt/Ti multilayers, which requires the quantification of the interfacial spin transparency. It is generally considered that there are two effects that can reduce $T_{int}$ below unity: spin backflow (SBF) due to the finite spin-mixing conductance of the interface [26-28] and spin memory loss (SML) due to interfacial spin-flip scattering [30]. According to the drift-diffusion analysis [26-30], the effect of SBF on $T_{int}$ is given by [35]

$T_{int}^{SBF} = [1-\text{sech}(d_{HM}/\lambda_s)][1+G_{HM}\tanh(d_{HM}/\lambda_s)/2G_{HM/FM}^{\uparrow\downarrow}]^{-1}$ (1)

where $\lambda_s$ and $G_{HM}=1/\rho_{xx}\lambda_s$ are the spin diffusion length and the spin conductance of the HM layer, and $G_{HM/FM}^{\uparrow\downarrow}$ is the bare spin mixing conductance of the HM/FM interface. $G_{Pt} \approx 1.3\times10^{15}$ $\Omega^{-1}$ m$^{-2}$ as determined by a thickness dependent spin-orbit torque experiment [35]. The theoretical value of $G_{Pt/Co}^{\uparrow\downarrow} = 0.59\times10^{15}$ $\Omega^{-1}$ m$^{-2}$ for the Pt/Co interface [27] is in reasonable agreement with the experimental values for Pt/FM interfaces [29,36] where the interfaces were engineered to reduce interfacial spin-orbit coupling and thereby minimize SML and two-magnon scattering [37]. Assuming a dominant Elliot-Yafet spin relaxation mechanism [33,34], we determined $\lambda_s$ and $T_{int}^{SBF}$ of the interface between the Co and Pt/Ti multilayers in Figs. 3(d) and 3(e), respectively. Due to the rapid increase in $\rho_{xx}$, $\lambda_s$ drops quickly from 2.9 nm for $m = 0$ (single Pt layer) to 0.6 nm for $m = 11$ (Fig. 3(d)). This places the multilayers in the bulk limit, $T_{int}^{SBF}= 0.5$, for $m \geq 3$ (see Fig. 3(e)). Recent work [30] has also established that the SML scales linearly with the interfacial perpendicular magnetic anisotropy energy density ($K_s$) of the HM/FM interface, which indicates the strength of ISOC. Specifically, $T_{int}^{SML} \approx 1-0.23K_s$ for the in-plane magnetized Pt/Co interface with $K_s$ in erg/cm$^2$. Our [Pt $d$/Ti 0.2]$_m$/Pt $d$/Co samples were deposited in a manner to minimize $K_s$ of the Pt/Co interfaces which we determined to range between 0.36 and 0.65 erg/cm$^2$ as $m$ was varied [38]. This indicated a relatively weak ISOC [30] and a SML which at the maximum would result in less than a 15% attenuation (see Fig. 3(e)).

From this determination of $T_{int} \approx T_{int}^{SBF}T_{int}^{SML}$, we obtain values for the internal $\theta_{SH}$ and $\sigma_{SH}$ for each different multilayer sample. As plotted in Figs. 3(a) and 3(b), $\theta_{SH}$ is enhanced from ~0.46 for the pure Pt sample ($m = 0$) to ~0.8 for $m = 7$ while $\sigma_{SH}$ is continually degraded by increasing $m$ (by a factor of ~3 for $m = 11$). Thus while $\theta_{SH}$ is increased by raising $\rho_{xx}$ of the spin Hall material the increase is much less than would be expected from a metal in the clean intrinsic limit where, if $\sigma_{SH}$ is constant, $\theta_{SH}$ should increase directly with $\rho_{xx}$.

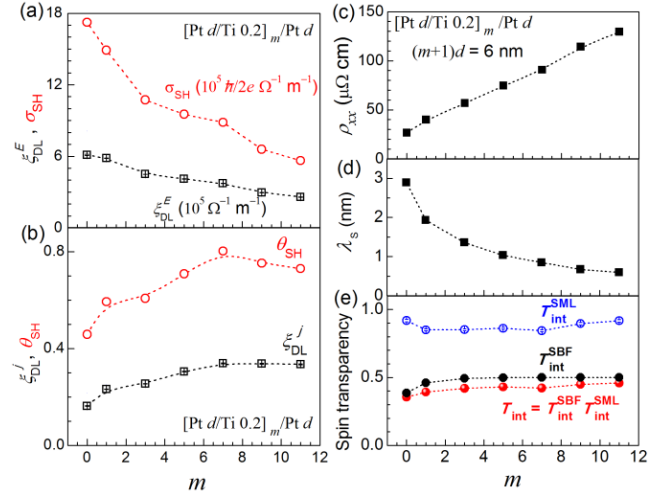

Fig. 3. (a) The dampinglike SOT efficiency per applied electric field $\xi_{DL}^{E}$ and the spin Hall conductivity $\sigma_{SH}$, (b) the dampinglike SOT efficiency per unit bias current density $\xi_{DL}^{j}$ and the spin Hall ratio $\theta_{SH}$, (c) the resistivity $\rho_{xx}$, (d) the spin diffusion length $\lambda_s$, and (d) the spin transparency for Pt/Ti multilayers with different $m$.

The possible mechanisms for the decrease of $\sigma_{SH}$ with decreasing $d$ (increasing $m$) are the key to understanding what determines the maximum attainable value of $\theta_{SH}$. We consider three effects that may affect the changes in the intrinsic $\sigma_{SH}$ as progressively more sub-monolayers of Ti are inserted into the Pt: degradation of the Pt crystal order, thin film strain that distorts the fcc symmetry and hence modifies the Pt band structure [39], and electron self-energy effects due to a shortened carrier lifetime that places the material into, or further into, the dirty-metal regime [19,21,23]. As revealed by the x-ray diffraction results in Figs. 4(a) and 4(b), there is a 50% reduction of the intensity of the Pt (111) diffraction peak ($I_{(111)}$) due to the increasing Ti insertions, which may be indicative of a moderate disruption of the fcc crystal order of Pt by the insertion layers. However, there seems to be no clear direct correlation between this degree of structural disruption with the degradation of $\sigma_{SH}$. In the case of the Pt/Ti multilayers studied here where Ti has a close atomic radius to Pt, $I_{(111)}$ decreases only by 50% while $\sigma_{SH}$ is reduced by 75%. This is in contrast to the result of a previous study of Pt/Hf multilayers [24], where $\sigma_{SH}$ was reduced only by 65% when $I_{(111)}$ was reduced by 95 %. With regard to possible strain effects, recent first-principles calculations have indicated that in-plane compressive strain can significantly reduce $\sigma_{SH}$ of Pt [40]. As shown in Fig. 4(c), the as-grown pure Pt samples are compressively strained in the film plane as indicated by the 0.6% increase of out-of-plane lattice plane spacing compared to that of the bulk Pt (111). We find that this strain is primarily the result of the low-pressure sputter deposition of the Pt onto the oxidized Si substrate, while the deposition of the Co overlayer adds a small additional contribution [38]. As indicated in Fig. 4(c), the in-plane compressive strain decreases with the increasing number of Ti insertions, while, in contrast, in the previous study of Hf insertions the in-plane compressive strain increased with increasing numbers of Hf layers [24]. These opposite changes in strain indicate that the similar decreases of $\sigma_{SH}$ that occur with both Ti or Hf insertions cannot be readily explained by strain distortion from the ideal fcc lattice structure. Unambigous evidence that the intrinsic spin Hall conductivity of Pt is rather robust against the moderate disorder (Fig. 4(b)) and the strain (Fig. 4(c)) is the essentially identical scaling of $\sigma_{SH}$ with the electrical conductivity $\sigma_{xx}$ for Ti insertions (black dots) and Hf insertions (red circles), which we plot in Fig. 4(d).

The exclusion of the degradation of the Pt crystal order and the effects of tensile strain leaves the shortening of the carrier lifetime [19,21,23] as the most likely mechanism for the decrease of $\sigma_{SH}$. As theoretically predicted [19,21], the shortening of the carrier lifetime should result in a rapid decrease of the intrinsic $\sigma_{SH}$ in the dirty-metal regime even if the crystalline order is fully maintained, with the transition point predicted as being $\rho_{xx} \approx 30$ μΩ cm for Pt. This appears to explain our experimental results rather well. For both Ti and Hf insertions (see Fig. 4(d)), $\sigma_{SH}$ decreases at first only gradually as the electrical conductivity $\sigma_{xx}$, which is proportional to the carrier lifetime, is decreased below the upper limit of the pure Pt film, and then decreases more rapidly with the decreasing $\sigma_{xx}$. This scaling of $\sigma_{SH}$ with $\sigma_{xx}$ is quite consistent with the predicted behavior in the dirty metal, short carrier lifetime, regime, unambiguously revealing that the shortening of carrier lifetime is the dominant mechanism of the reduction of $\sigma_{SH}$ in the low $\sigma_{xx}$ (high $\rho_{xx}$) regime.

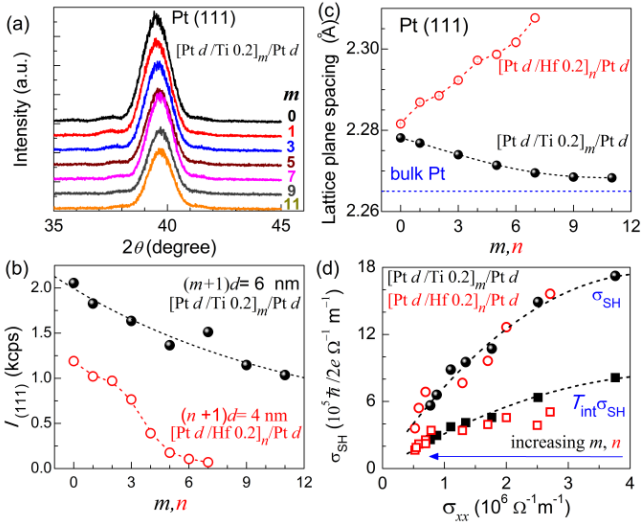

Fig. 4. (a) X-ray diffraction $\theta$-$2\theta$ patterns for [Pt $d$/Ti 0.2]$_m$/Pt $d$. (b) The integrated intensity of Pt (111) peak $I_{(111)}$ and (c) Lattice plane spacing of Pt (111) for [Pt $d$/Ti 0.2]$_m$/Pt $d$ and for [Pt $d$/Hf 0.2]$_n$/Pt $d$, with total Pt thickness $d(m+1) \approx$ 6 nm for Pt/Ti multilayers and $d(n+1) \approx$ 4 nm for Pt/Hf multilayers. (d) Scaling of the spin Hall conductivity with electrical conductivity for [Pt $d$/Ti 0.2]$_m$/Pt $d$ (solid, black) and for [Pt $d$/Hf 0.2]$_n$/Pt $d$ (red, open), indicating the dominating role of carrier lifetime in degradation of spin Hall conductivity. The dashed lines in (c)-(d) are to guide the eyes.

As indicated theoretically [19,21,23], the high-resistivity dirty metal behavior of the spin Hall conductivity of Pt sets the limit to which $\theta_{SH} = \sigma_{SH}/\sigma_{xx}$ ($\xi_{DL}^j = T_{int}\sigma_{SH}/\sigma_{xx}$) can be enhanced via increasing the resistivity of Pt based spin Hall material. Since this reduction of $\sigma_{SH}$ cannot be avoided in any process that shortens the carrier lifetime, it sets a limit to which $\theta_{SH}$ ($\xi_{DL}^j$) of Pt can be enhanced by increasing the scattering rate. Although the exact theoretical limit of $\theta_{SH}$ and the corresponding resistivity have remained unsettled, the maximum value of $\theta_{SH} \approx 0.8$ that we have obtained for [Pt 0.75/Ti 0.2]$_7$/Pt 0.75 indicates an upper bound for $\theta_{SH}$ of at least 0.8 for the intrinsic SHE of Pt. The corresponding maximum $\xi_{DL}^j$ is $\geq 0.4$ ($T_{int}^{SBF}$=0.5, $T_{int}^{SML}$=1).

Figure 1(b) compares the experimental values of $\xi_{DL}^j$ and $\rho_{xx}$ for various strong spin Hall metals. $\xi_{DL}^j = 0.34$ ($\rho_{xx} = 90$ μΩ cm) for the optimum Pt/Ti multilayers is comparable to that of optimum Pt/Hf multilayers (144 μΩ cm)[24] and the best values for $\beta$-W ($\rho_{xx} = 200$-300 μΩ cm)[8,13,14], whereas it is higher than that for less-resistive Pt (20-50 μΩ cm) [7,25], Pd$_{0.25}$Pt$_{0.75}$ (57 μΩ cm) [12], Au$_{0.25}$Pt$_{0.75}$ (83 μΩ cm)[11], Pt$_{0.6}$(MgO)$_{0.4}$ (74 μΩ cm) [23], and more resistive Pt$_{0.85}$Hf$_{0.15}$ [22], $\beta$-Ta [2]. As we calculated in Fig. S3 within the Supplementary information, SOT-MRAMs based on the optimum Pt/Ti multilayers is also much more energy-efficient than the conventional spin Hall metals (Pt [25], $\beta$-Ta [2], and $\beta$-W [13]) and the topological insulators (Bi$_x$Se$_{1-x}$ [16]). Therefore, the multilayer [Pt 0.75/Ti 0.2]$_7$/Pt 0.75 with giant $\xi_{DL}^j$ and relatively low $\rho_{xx}$ at the same time is a very compelling spin Hall material for spin-torque applications (e.g. memories, oscillators, and logic) that require high current/energy efficiency ($I_{write} \propto (1+s)/\xi_{DL}^j$, $P_{write} \propto [(1+s)/\xi_{DL}^j]^2\rho_{xx}$), low device impedance ($\propto \rho_{xx}$) and high endurance.

In conclusion, we have presented a systematical study on the evolution of the SHE and $\rho_{xx}$ of Pt with interfacial scattering from sub-monolayer Ti insertion layers. We obtained an approximate doubling of the spin torque efficiency $\xi_{DL}^j$ when $\rho_{xx}$ was increased from 26.5 to 130 μΩ cm via the strong interfacial scattering that was the result of the insertion of increasing numbers of sub-monolayers of Ti. At the same time there was a factor of 3 reduction of the spin Hall conductivity $\sigma_{SH}$ of the material. We have concluded that this reduction of $\sigma_{SH}$ is mainly due to the effect of the decrease of carrier lifetime, whereas $\sigma_{SH}$ of the fcc Pt is insensitive to the strain and the moderate interruption of crystal order caused by the Ti sub-monolayer insertions. Since the reduction of $\sigma_{SH}$ with shortening carrier lifetime is an inherent characteristic of the intrinsic SHE of Pt, it sets an effective upper bound of $\xi_{DL}^j \geq 0.4$ ($\theta_{SH} \geq 0.8$) for materials where Berry curvature of the Pt band structure is the source of the SHE. This work also establishes a highly-efficient spin-current generator, [Pt 0.75/Ti 0.2]$_7$/Pt 0.75, that combines the maximum $\xi_{DL}^j$ with a relatively low $\rho_{xx}$ (90 μΩ cm), and good compatibility with integration technology (e.g., simple growth with standard sputtering techniques on Si substrates) for development of low-power, low-impedance, and high-endurance magnetic memories, oscillators, and logic.

**Acknowledgments**
This work was supported in part by the Office of Naval Research (N00014-15-1-2449), and by the NSF MRSEC program (DMR-1719875) through the Cornell Center for Materials Research. This work was performed in part at the Cornell NanoScale Facility, an NNCI member supported by NSF Grant ECCS-1542081.